# New investigation of the electronic and structural properties of (Mg,Ti)-doped and co-doped ZnO structures: A DFT and DFT+U study

Sidi Ahmedbowba[1], Fehmi Khadri[1], Walid Ouerghui[2], Said Ridene[3, *]

[1]*LSAMA, Phys. Dept, University of Tunis El Manar, Tunis, 2092, Tunisia.*

[2]*LPCM, Phys. Dept, University of Tunis El Manar, Tunis, 2092, Tunisia*

[3]*Advanced Materials and Quantum Phenomena Laboratory, Physics Department, Faculty of Sciences of Tunis, Tunis El-Manar University, 2092 Tunis, Tunisia.*

* Corresponding author. E-mail Address: said.ridene@fst.rnu.tn

**Abstract**

This study investigates the novelty of the crystalline and electronic structure of (Mg,Ti)-doped ZnO and the co-doped $Zn_{1-x-y}Mg_xTi_yO$ structures using Gaussian and plane-wave basis sets, as implemented in the CP2K code. The goal of incorporating low concentrations of Mg and Ti into ZnO is to influence its electronic properties without significantly altering its geometrical and crystalline structure. Within the framework of density functional theory (DFT), we analyze various doped and co-doped configurations. Our results show that Ti-doped ZnO exhibits an indirect band gap, while Mg doping preserves the direct semiconductor behavior of ZnO structure, with an increase in band gap energy. Additionally, the co-doped $Zn_{1-x-y}Mg_xTi_yO$ system, at varying concentrations of Ti and Mg, displays minimal lattice deformation. These findings suggest that this material could be a promising candidate for transparent electronic devices, highlighting the importance of understanding the electronic structure of ZnO to optimize its physical properties.

## 1. Introduction

Recent research has shown a growing interest in the utilization of Zinc oxide structure (ZnO) as a transparent conducting oxide (*TCO*) due to its exceptional optical transparency and low electrical resistivity [1-3]. With a wide band gap of 3.4 eV [4], high breakdown voltage and a significant exciton binding energy of 60 meV [5,6], it exhibits a range of valuable properties for semiconductor applications such as flat screens, solar cells, and smart coatings. The ZnO can crystallize in three different structures which are the Zinc blend, Rocksalt and the Wurtzite one. As shown by *ab initio* studies [7], the last structure is the most energetically stable one. The availability of ZnO in this structure in the form of single crystals and large-area substrates further enhances its appeal for a variety of technological advances. To take advantage of these characteristics, doping processes are often used. For example, as some studies have shown [8], doping ZnO with magnesium (Mg) can increase its bandgap energy. Indeed, the energy of the fundamental band gap of $(Zn_{1-x}Mg_x)$ can vary between 3.4 eV and 7.6 eV, depending on the Mg ratio. Since zinc and magnesium have similar ionic radii (0.60 $a_0$ for $Zn^{2+}$ and 0.57 $a_0$ for $Mg^{2+}$) and the same number of fourfold coordination, the alloy can be formed by replacing a Zn atom with an Mg atom without modifying the original structure. Thus, due to its partially filled sub-shells, Zinc host ions can be easily substituted by ions of Mg or another compound with similar properties [9]. As a result, the transition metal Ti proves to be a highly suitable compound by comparison with other materials. In fact, Ti is a non-ferromagnetic element with a higher ionic radius than Zn [10]. It has one more valence electron and allows for a shorter bond length. The properties of ZnO could be easily modified and controlled by Ti doping [11], as it emerges as a potential alternative dopant for improving mobility and electrical conductivity in ZnO crystals [12].

By doping and co-doping ZnO with varying concentrations of Mg and Ti as donor atoms, this study aims to explore the optimal configuration for achieving desired material properties. Note that the investigation of the electronic and structural properties of the incorporation of Mg and Ti in ZnO structure and the detailed experimental and theoretical studies are still lacking. However, through the density functional theory (DFT) calculations and a focus on determining the ideal concentrations of Mg and Ti for incorporation, this work seeks to pave the way for the development of *n*-type compounds with enhanced characteristics [10]. We have doped ZnO with three different concentrations (2.77 %, 5.55 %, and 8.33 %) of Mg or Ti in various positions. Finally, we co-doped ZnO with both Ti and Mg at concentrations of 5.54 %, 11.10 %, and 16.66 % to investigate the effects of these dopants on the material's properties. These atoms were placed in various sites occupied by Zn in the supercell of ZnO to find the most energetically favorable configuration, which was then used to calculate some physical properties. Currently, DFT calculations face challenges related to the band-gap issue prevalent in many transition metal oxides such as ZnO. Specifically, the local density approximation (LDA) and the generalized gradient approximation (GGA) [13] tend to underestimate the band gap value when compared to its experimental value of 3.4 eV. To address this discrepancy, several modifications have been introduced, such as the LDA+U and GGA+U methods, which incorporate the Hubbard U parameter to enhance the accuracy of material property descriptions in the LDA or GGA calculation [14]. Additionally, the Heyd-Scuseria-Ernzerhof (HSE) hybrid functional [15] and the GW approximation [16] have been considered to further refine these calculations.

## 2. Computational methods

In this study, we used the CP2K code [17] to investigate the impact of Ti and Mg substitution sites within the ZnO supercell. A wurtzite supercell of ZnO was constructed with 3x3x2 dimensions, comprising a total of 72 atoms. Periodic boundary conditions were applied along the X, Y, and Z directions to accurately capture the local environment and electronic interactions of the dopants. Goedecker-Teter-Hutter (GTH) pseudopotentials were used to describe the core electrons, while the valence electrons were treated explicitly with double-zeta valence polarized (DZVP) basis sets, allowing a balance between calculation cost and accuracy [18]. Several exchange-correlation functionals within the GGA (generalized gradient) approach were tested to ensure the reliability of our results. Among these functionals, the PBE (Perdew-Burke-Ernzerhof) one was used due to its proven effectiveness in describing the electronic structure of semiconductors and insulators. Additionally, geometry optimizations were performed using the BFGS (Broyden-Fletcher-Goldfarb-Shanno) algorithm, with convergence criteria for the maximum force and energy change set to $10^{-4}$ Hartree/$a_0$ and $10^{-6}$ Hartree, respectively. The electronic structure calculations were conducted using the Kohn-Sham formulation of density functional theory (DFT). The plane-wave cut off energy was set to 600 Ry, and a Monkhorst-Pack k-point mesh of 3x3x2 was used to sample the Brillouin zone. So, the $KPOINT_{SET}$ was carefully chosen to ensure convergence of the electronic properties, balancing computational efficiency and accuracy. As shown by several computational studies using LDA and GGA approximations, density functionals generally underestimates band gaps for semiconductors and sometimes incorrectly predicts a metal behavior [19-21], as indicated in Table 2. The solution for this calculation error has long been discussed and methods such as the scissor operator [22], the self-interaction correction (SIC) [23,24] or the Hubbard-U correction [25] have been suggested to increase the band gap estimation. Among these corrections, the Hubbard-U method [26] is widely used due to its simple calculation procedure, reliable results, and practical efficiency. This method introduced an orbital-dependent term known as on-site Coulomb repulsion energy U, into the XC term of the LDA and GGA approaches. They are now referred as LDA+U or GGA+U and expressed as follows [21]:

$$E_{LDA+U}[n(r)] = E_{LDA}[n(r)] + E_U[n(r)] - E_{dc}$$

where, $n(r)$ is the electron density, $E_{LDA}$ is the energy from the conventional LDA functional, $E_U$ is the Hubbard type energy, and $E_{dc}$ is the double-counting correction energy. The on-site Coulomb interactions are particularly significant for localized $d$ and $f$ electrons but can also be important for $p$-localized orbitals [27,28]. Hence, LDA+U or GGA+U has potentially improved the insufficient description of strongly localized electrons, such as those in the Zn-$d$ state, which are not correctly described by LDA and GGA methods. In this work, we tested $U_{d\text{-}Zn}$=10 eV and $U_{p\text{-}O}$=7 eV since these Hubbard-U correction values have been used in other theoretical studies [21] and yield gap energy values close to experimental results, as shown in Table 2.

Using this correction which is implemented in the CP2K code only for the high symmetry k-point Γ, we performed geometry optimization, LUMO (Lowest Unoccupied Molecular Orbital) and HOMO (Highest Occupied Molecular Orbital) estimation and calculated the band gap values at the Γ k-point. We also generated the Projected Density of States (PDOS) for each element along with the total Density of States (DOS). Since the DFT+U method is not yet implemented in CP2K for calculation along high symmetry k-points (other than Γ), we used a Python code to combine the band structure calculated through GGA with the LUMO-HOMO, band gap values from DFT+U calculation, to generate a correction factor for the band structure. In order to validate our computational approach and to confirm the reliability of our chosen methods and functionals, we benchmarked the computed lattice parameters, band gaps and formation energies against available experimental data and previous theoretical studies. The aim of all these calculations is to investigate the electronic and structural properties of ZnO doped and co-doped with Ti and Mg.

## 3. Results and Discussion

### 3.1. Structural and electronic properties optimization of pure ZnO

The optimization of structural properties is crucial for theoretical calculations and the determination of other physical properties. In this study, we focused on calculating the total energy $E_{tot}$ of various compounds, including the pure ZnO. This optimization was performed using the BFGS minimizer implemented in the CP2K code, which allowed us to achieve highly accurate results. As the main aim of this work is the study of doping and co-doping, the study of the pure lattice is only carried out to check that we are in the same order of magnitude as that obtained experimentally. Thus, after geometric optimization, the comparison with previous studies presented in Table 1, show that our results for the pure ZnO lattice are in good agreement with the experimental one, with a relative error (given in brackets in the table) of no more than 3% for d1ZnO (distance between the Zn atom and its nearest O atom neighbor), 1.5% for d2ZnO (distance between the Zn atom and its second nearest O atom neighbor) and 0.7 % for the other lattice parameters (a and c). This highlights the quality and reliability of our results which align well with the experimental values. We should note here that the optimized configuration (panel (a) of figure 1) was used as backbone for all calculations.

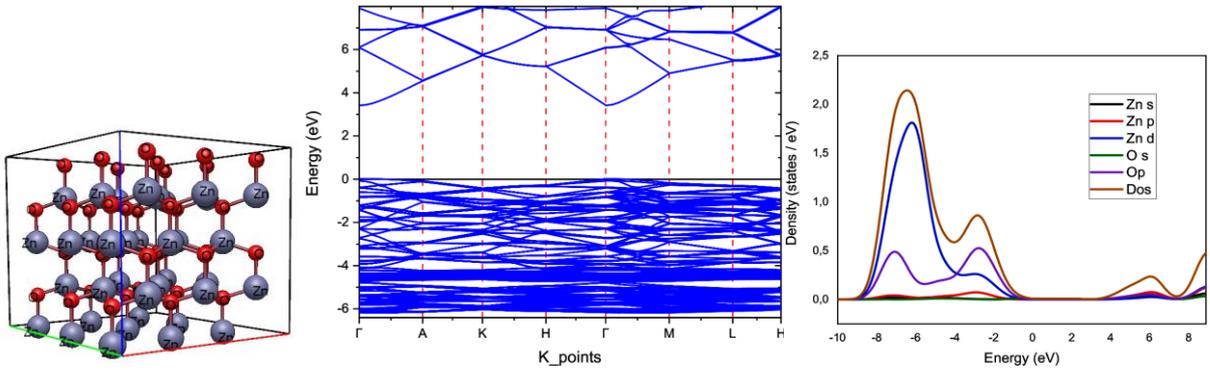

**Fig. 1.** *Optimized configuration (a) (panel a), band structure (panel b) and, the PDOS and DOS (panel c) of the (3x 3x2) pure ZnO supercell.*

*Table 1: Comparison of ZnO cell parameters with earlier theoretical and experimental ones.*

| Method | a (Å) | c (Å) | c/a | d$_{1ZnO}$(Å) | d$_{2ZnO}$(Å) | $E_{tot}$ (a.u) |
|---|---|---|---|---|---|---|
| *This work* | 3.2576(0.64%) | 5.2047(0.33%) | 1.6137(0.05%) | 2.0073(2.87%) | 2.0073(1.42%) | -2757.3383 |
| *Theory*[3,29] | 3.2489(0.35%) | 5.2066(0.29%) | 1.6026(0.64%) | 1.9876(1.85%) | 1.9875(0.45%) | - |
| Exp[30,31] | 3.2375(0.%) | 5.2220 | 1.6129 | 1.9512 | 1.9791 | - |

In order to investigate the ZnO electronic properties, as it behaves as a semiconductor, we performed band structure calculations of pure ZnO along high symmetry k-points: $\Gamma$(0 0 0), $A$(0 0 1/2), $H$(1/3 2/3 1/2), $K$(1/3 2/3 0), $M$(1/2 0 0), and $L$(0 1/2 1/2). Thus, we derived the band gap energy ($E_g$) which is presented in Table 2. Using the *GGA* approximation, we found a band gap energy value of 0.71 eV, which is close to other theoretical studies using DFT. This value is 79 % less than the experimental one, which is 3.4 $Ev$ [4]. To avoid this underestimation, we employed the DFT+U method, which gave a band gap energy of Eg=3.401eV. The relative error with respect to experimental values shows the great improvement in calculation accuracy achieved by adopting this correction, with an error of no more than 0.03 %. We also used the same method to generate the Projected Density of States (PDOS) for each element (Zn and O) along with the total density of states (DOS). The band structure calculations, presented in panel b of figure 1, indicate that ZnO has a direct band gap at the Gamma k-point. It is a characteristic that is beneficial for optical applications as it allows for efficient light absorption and emission. This direct band gap nature is essential for enhancing the performance of optoelectronic devices, such as light-emitting diodes (LEDs) and laser diodes. The band gap energy found in this study is 3.401 eV, which is close to both theoretical and experimental values reported in the literature. In fact, we observe that the calculated band gap energy in this work (3.401 eV) aligns well with both theoretical (3.380 eV) [32] and experimental values (3.440 eV) [33]. The observed slight deviation is within the expected range of computational accuracy and experimental measurement tolerances. This confirms the semiconductor nature of ZnO, which is essential for its functionality in electronic and optoelectronic devices.

*Table 2: Theoretical and experimental band gap energies of ZnO using different methods.*

| Method | Gap Energy Eg (eV) |
|---|---|
| Theory DFT [34] | 0.794(76.59%) |
| This Work DFT | 0.710 (79.12%) |
| Theory DFT+U [35] | 3.370 (0.88%) |
| This Work DFT+U | 3.401$ (0.03%) |
| Spectroscopy UV- VIS [36] | ≈3.4 |
| Spectroscopy X-ray [37] | ≈3.3 |

To illustrate the contributions of each electron orbital to the energy band, we have exhibited the pure ZnO DOS together with the PDOS of both zinc and oxygen in panel c of Figure 1. The DOS and PDOS plots cover the energy range from -10 eV to 9 eV, with the Fermi level set to 0 eV. The calculated band gap value (3.401 eV) is indicated in the DOS plot. The two zones of the valence band, ranging from -4 eV to 0 eV and -8 eV to -4 eV, are primarily derived from the O(p) state and Zn(d) state, respectively. The energy band at the bottom of the conduction band is mainly composed of the Zn(4p) state, accompanied by a significant transition process between Zn(4p) and O(2p) states. This detailed analysis of the DOS and PDOS highlights the significant contributions of the different atomic states to the electronic structure of ZnO, providing deeper insights into its electronic properties and potential applications. These electronic properties analysis confirms that ZnO's band structure and gap energy are consistent with its known semiconductor properties, supporting its widespread used in various advanced technologies. However, adding some dopants can increase these electronic features.

### 3.2. Doping with Magnesium (Mg)

### 3.2.1 Structural Properties Optimization

After studying the pure ZnO cell containing 72 atoms, we proceeded to introduce Magnesium (Mg) as a doping agent by replacing a Zinc atom with Mg at a concentration of 2.77% as shown in Figure 2.

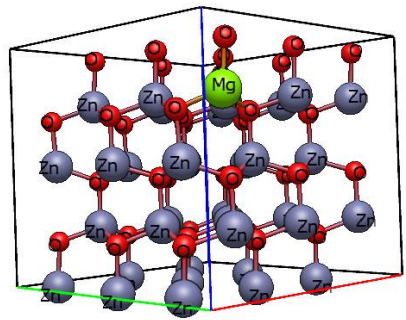

*Fig. 2. Configuration (b)of ZnO dropped by 2.77% Mg.*

We compared the results of the doped cell with $2.77$\% Mg to those of the pure cell found in this work and other experimental data studies. We focused on parameters such as lattice constants (a and c), distances between atoms, including Zn-O, Zn-Mg, and O-Mg, as listed in Table 3.

*Table 3. Structural parameters for ZnO doped by 2.77 % Mg.*

| Parameter | ZnO | Mg-ZnO | Mg-ZnO (Exp) |
|---|---|---|---|
| a (Å) | 3.2375(0.51%) | 3.2489(0.84%) | 3.221 [38] |
| c (Å) | 5.2220(0.61%) | 5.2066(0.02%) | 5.191 [38] |
| c/a | 1.6129(1.15%) | 1.6026(0.55%) | 1.5948 [38] |
| $d_{1ZnO}$ (Å) | 1.9512(0.71%) | 1.9950(0.90%) | 1.97706 [39] |
| $d_{2ZnO}$ (Å) | 1.9791 | 2.0000 | - |
| $d_{ZnMg}$ (Å) | -- | 2.0250 | - |
| $d_{MgO}$ (Å) | -- | 3.2444 | - |
| $E_{tot}$ (a.u) | -2757.3383 | -2763.0281 | - |

With this type of spiking, our calculations give an error relative to experience of no more than 1\%, giving an idea of the accuracy of our calculations for this system. This gives us confidence in the quality of the results to be obtained on electrical properties such as band structure calculations. From Table 3, we have extracted some interesting key points. However, the lattice constant *a* increased slightly from 3.2375 Å to 3.2489 Å, while the lattice constant *c* decreased from 5.2220 Å to 5.2066 Å. This minor expansion in the *a*-axis and contraction in the c-axis can be attributed to the incorporation of $Mg^{2+}$ ions, which have a slightly smaller ionic radius compared to $Zn^{2+}$ ions. These changes in lattice constants are consistent with findings from other studies, which report that Mg doping typically results in a slight distortion of the ZnO lattice without significantly altering its wurtzite structure [39, 40]. Thus, the ratio c/a decreased from 1.6129 to 1.6026, indicating a small anisotropic change in the lattice parameters. This result suggests that Mg doping induces a subtle change in the ZnO lattice symmetry, which can affect the material's electronic properties. Similar trends in the ( c/a ) ratio have been observed in other doped ZnO systems [25]. On the other hand, the Zn-O bond distances, $d_{1ZnO}$ and $d_{2ZnO}$, increased slightly from 1.9512 Å and 1.9791 Å to 1.9950 Å and 2.0000 Å, respectively. This elongation of Zn-O bonds upon Mg doping is consistent with the lattice expansion in the a-axis direction. The introduction of Mg resulted in new bond distances, such as $d_{ZnMg}$ (2.0250 Å) and $d_{MgO}$ (3.2444 Å). The $d_{ZnMg}$ distance indicates that the Mg atoms are slightly away from the Zn atoms compared to the Zn-O bonds, while the $d_{MgO}$ distance shows a larger separation, which is typical due to the lower electronegativity of Mg compared to Zn. Therefore, the total energy of the supercell decreased by 0.2 % as impact of presence of 2.77 % of Mg. Overall, the structural parameters observed in this study align well with previously reported values for Mg-doped ZnO [38-40]. The slight changes in lattice constants and bond lengths are indicative of successful doping, which minimally perturbs the host lattice while potentially altering its electronic properties.

To explore this further, we increased the Mg concentration in the cell in different positions. Thus, we optimized three different cells (depicted in Figure 3) to study the effects of Mg concentration and position on parameters like lattice constants and inter-atomic distances.

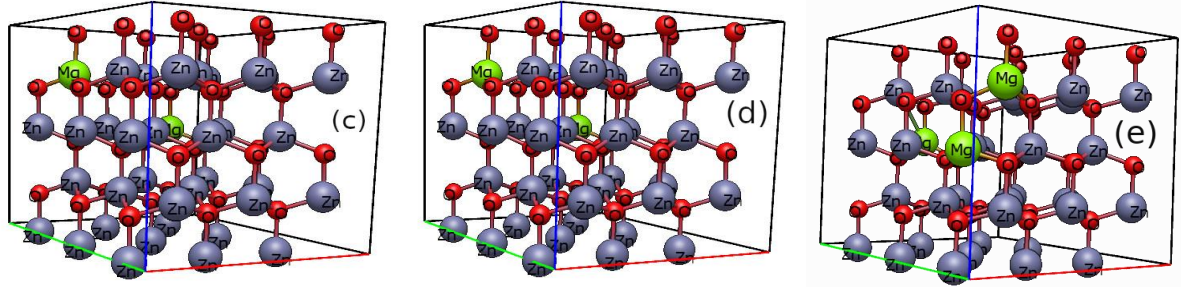

Fig. 3. Configurations (c), (d) and (e) of ZnO doped with 5.55%, 5.55%, and 8.33% of Mg concentration respectively.

These parameters presented in Table 4 show that configuration (c) and configuration (d), each doped with two Mg atoms representing 5.55%, have minimal variation in the interatomic distances compared to the undoped ZnO. Configuration (e) doped with three Mg atoms which correspond to 8.33%, exhibit no significant changes in the interatomic distances which are close to configuration (c). This is promising, unless doping with higher Mg concentration doesn't change the gobal supercell configuration, but with a decrease in the total energy. By comparison with the pure ZnO cell, the total energy $E_{tot}$ of configurations (b), (c), (d) and (e) increase with the Mg concentration but the position doesn't have significant impact on it. Subsequently, these optimized configurations were used to perform band structure calculations to investigate the electronic properties of ZnO doped with Mg.

Table 4. Interatomic distances and total energy for different configurations of $Zn_{1-x}Mg_xO$.

| Configuration | d1Zn-O (Å) | d2Zn-O (Å) | dMg-O (Å) | dMg-Zn (Å) | $E_{tot}$ (a.u) |
|---|---|---|---|---|---|
| c | 2.000 | 1.998 | 2.007 | 3.110 | -2763.1822 |
| d | 1.999 | 1.994 | 2.025 | 3.243 | -2763.1761 |
| e | 2.003 | 2.018 | 2.018 | 3.193 | -2765.9712 |

### 3.2.2 Electronic Properties

To analyze the electronic properties of ZnO doped with Mg in configurations (b), (c), (d), and (e), we performed a band structure calculation along high symmetry k-points: Γ(0 0 0), A (0 0 1/2), H (1/3 2/3 1/2), K (1/3 2/3 V0), M (1/2 0 0), and L (0 1/2 1/2). These band structures are depicted in Figure 4. The calculated band gap energies for each configuration were found to be 3.37 eV, 3.48 eV, 3.50 eV, and 3.56 eV, respectively.

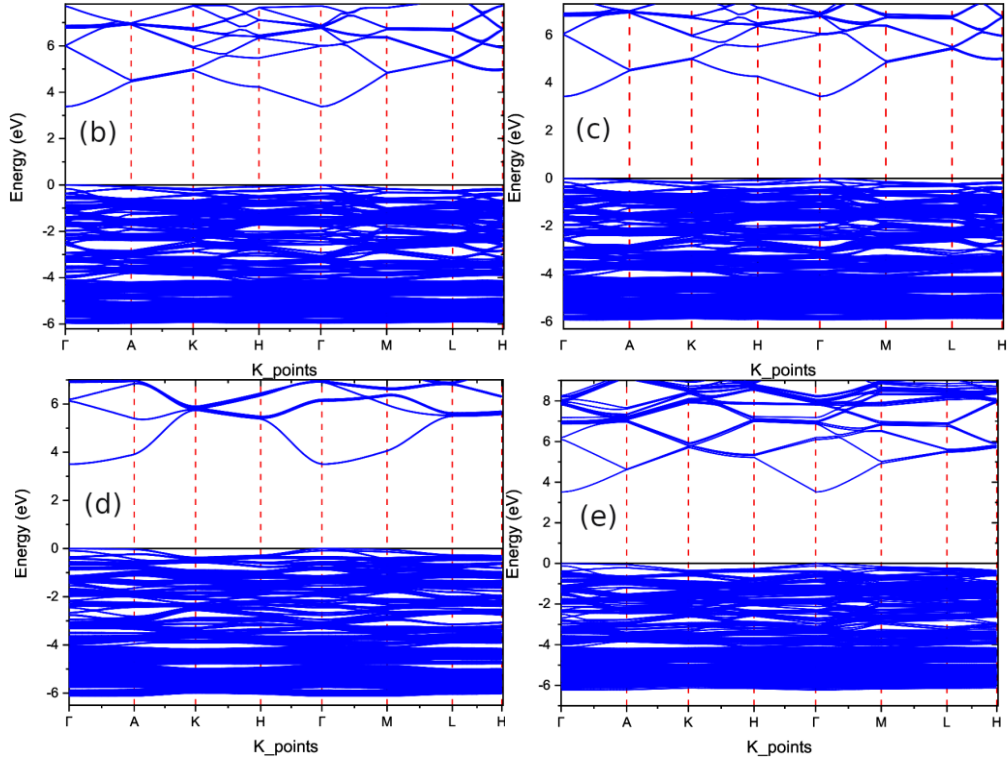

*Fig. 4. Band structure of ZnO doped for different concentrations (2.77%, 5.55%, 5.55%, and 8.33%) and for different configuration (b),(c),(d) and (e) of Mg. The Fermi level has been referenced to 0.*

The band gap values obtained were consistent with experimental data, confirming the effectiveness of our doping strategies in modifying the electronic structure of ZnO. Indeed, in configuration (b) where Mg atoms are strategically positioned to assess their impact on ZnO's electronic structure, the band gap of 3.37 eV indicates a slight narrowing compared to that of pure ZnO. This suggests the presence of localized states near the valence band edge due to Mg incorporation. Configuration (c), doped with two Mg atoms (5.55%), exhibits a band gap of 3.48 eV, showing a moderate alteration in electronic behavior. Similarly, configuration (d), also doped with two Mg atoms but in different positions, demonstrates a band gap of 3.50 eV, indicating subtle variations in electronic states possibly due to different Mg-Zn interactions. Notably, configuration (e), with three Mg atoms (8.33%), shows a band gap of 3.56 eV which is a wider value than that of the other configurations and suggests further localization effects or changes in the band structure induced by higher Mg concentrations. For all studied configurations, the band gap was found to be direct at the Γ k-point, emphasizing the direct nature of electronic transitions within the ZnO band structure.

*Table 5: Band gap energies (Eg), in eV, of ZnO configurations doped with Mg compared to the available experimental data.*

| Configuration | This Work | | Experimental |
|---|---|---|---|
| | DFT | DFT+U | |
| (b) | 0.753(77.71%) | 3.370 (0.30%) | 3.380 [38] |
| (c) | 0.871(74.65%) | 3.480(1.46%) | 3.430 [38] |
| (d) | 0.881(74.38%) | 3.501 (2.07%) | 3.430 [38] |
| (e) | 0.902 | 3.567 | - |

As we saw in the calculation of the pure cell, the U correction greatly improves the accuracy of the calculations, and this is confirmed by the values obtained for ZnO doped with Mg. These values, of the band gap energies (*E*g) corresponding to different configurations, presented in Table 5 and compared with the experimental ones available in the literature, can tell us which is the most realistic structural arrangement. Indeed, configuration (b) seems to be the closest to experiment, with a relative error of 0.3%, so it's probably the most possible to actually achieve. For the 2.77% doped configuration, the calculated band gap of 3.37 eV shows a slight deviation from the experimental value of 3.380 eV [38]. The doped configuration in this example is particularly interesting because it improves the electronic properties of ZnO doped with Mg, presenting a new challenge for experimenters to find the right technologies to achieve these configurations. Configurations (c) and (d), corresponding to 5.55% Mg doping, exhibit calculated band gaps (3.48 eV and 3.50 eV, respectively) both in close agreement with the experimental value of 3.43 eV [38]. Once again, the double Mg doping (with two Mg atoms) maintains the same electronic properties as pure ZnO, confirming the great potential of this $Mg_x Zn_{1-x}O$ compound. Despite the absence of experimental data, the configuration (e), corresponding to 8.33% Mg doping, shows a calculated band gap (3.56 eV) close to that of the 5.55% configuration. We can suggest that the latter is more suitable for achieving a good compromise between improved electronic properties and maintaining good structural stability of our configuration. To illustrate contributions of each electron orbital to the band structure, we have exhibited in figure 5 the total density of states (DOS) of the concentration (e) together with the Projected Density of States (PDOS) of zinc, Magnesium, and oxygen.

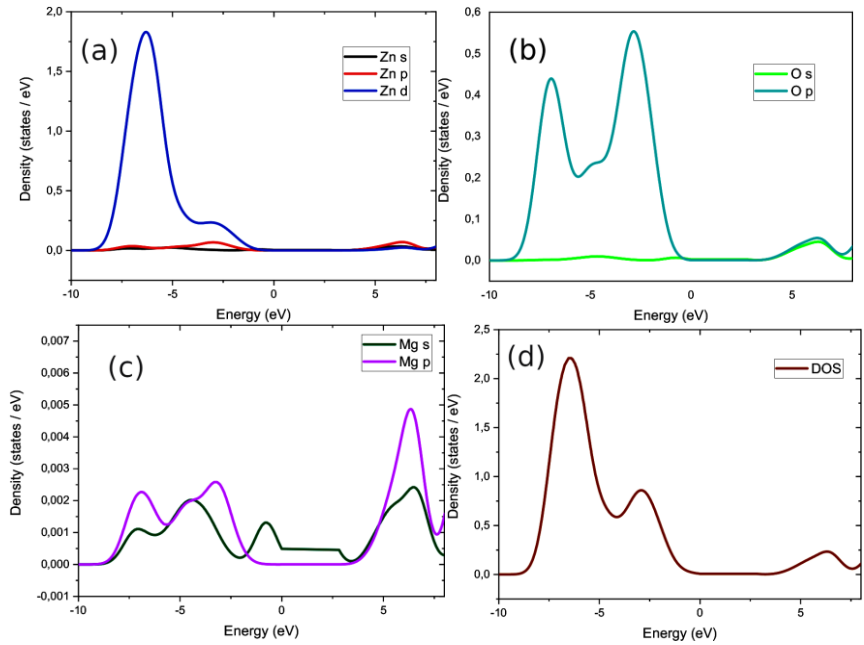

*Fig. 5. DOS and PDOS for the configuration (e). Panels (a), (b), and (c) show the contributions from Zn, O, and Mg atoms respectively, while panel (d) displays the total density of states. The Fermi level is referenced to 0 eV.*

The DOS and PDOS plots cover the energy range from -10 eV to 8 eV, with the Fermi level set to 0 eV. The calculated band gap of 3.56 eV is indicated in the DOS plot. The two zones of the valence band, ranging from -4 eV to 0 eV and -8 eV to -4 eV, are primarily derived from the Zn(d) state (panel (a) of figure 5 and O(2p) state (panel (b)), respectively. The energy band at the bottom of the conduction band is mainly composed of the Zn(4p) state, accompanied by a significant transition process between Zn(4p) and O(2p) states. Additionally, a small density of about 0.002 at the end of the valence band is derived from the Mg(s) state, and a small density of 0.005 at the beginning of the conduction band is derived from the Mg(p) state. This explains the difference in the band gap between the pure supercell and that doped with 8.33 % of Mg. These findings underscore the sensitivity of ZnO's electronic structure to Mg doping configurations, demonstrating the potential for tailored electronic properties through controlled doping strategies.

### 3.3. Doping with Titanium (Ti)

### 3.3.1 Structural Properties Optimization

After investigating the pure ZnO cell containing 72 atoms and that one doped with Mg, we introduced Titanium (Ti) as another doping alternative such as Mg. This is achieved by substituting Zinc atoms with Ti at different concentrations (2.77 %, 5.55 % and 8.33 %). The ZnO doped with a concentration of 2.7 % is shown in Figure 6. We should note here that this configuration is obtained after geometric optimization. In the same way, Table 6 provides comprehensive data on the structural properties and interatomic distances of ZnO doped with Ti.

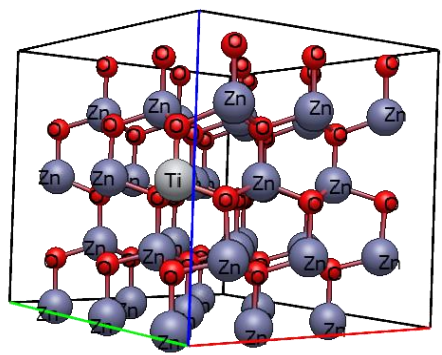

*Fig.* 6. Configuration (f) of ZnO doped with 2.77 % Ti.

The analysis of the main findings is as follows: For the lattice parameters (a, c and c/a), we notice that the parameters *a* and *c* show an increase in Zn-TiO compared to the pure ZnO, indicating an expansion of the crystal structure upon Ti doping. While the ratio c/a suggests a more compact structure along the c-axis relative to the a-axis in the Zn-TiO Cristal. Those findings were reported by other studies and may have an impact on other properties of the compound [41,42]. For the interatomic distances (d1ZnO, d2ZnO, dZnTi and dTiO), we notice that the distances d1ZnO and d2ZnO between Zn and O atoms show an increase upon Ti doping, since Zinc and Titanium have different ionic radii (0.60 Å for $Zn^{2+}$ and 0.68 Å for $Ti^{2+}$) [9,44]. For the total energy we note an increase of ~59 eV, indicating a reduction of the stability of Zn-TiO. To further explore the impact of Ti atoms on the cell, we increased the Ti concentration in the supercell and changed its position to assess the impact on ZnO properties. So, we optimized three different cells depicted in Figure 7.

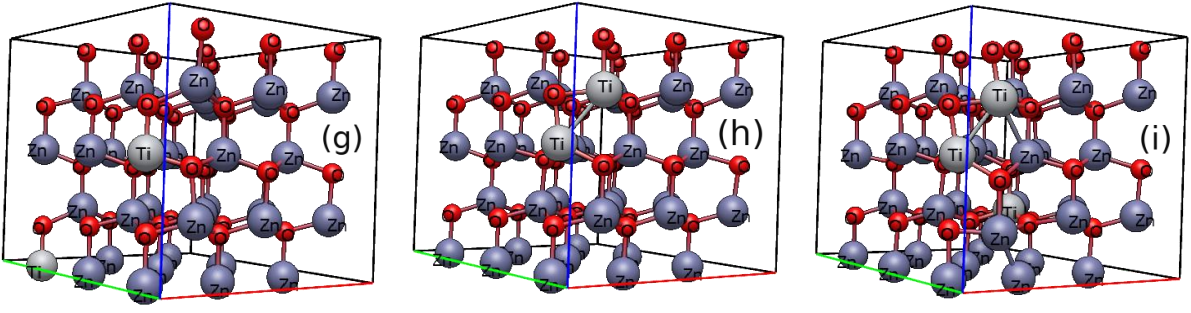

*Fig. 7. Configurations (g), (h) and (i) of ZnO doped with $5.55\%$, $5.55\%$ and $8.33\%$ of Ti, respectively.*

The obtained parameters were given in Table 6, where we observe that configuration (g) and configuration (h), each doped with two Ti atoms representing $5.55\%$ doping, have minimal variation in the interatomic distances compared to the ZnO configuration (i). This one doped with three Ti atoms (8.33%), exhibited slightly different bond lengths, particularly in dTi-O and dTi-Zn, suggesting a nuanced impact of higher Ti concentration on the local structure.

*Table 6: ZnO in Wurtzite geometry (72 atoms) doped with 2.77% Ti.*

| Parameter | ZnO | Zn-TiO |
| --- | --- | --- |
| a (Å) | *3.2375* | *3.4109* |
| c (Å) | 5.2220 | 5.2266 |
| c/a | 1.6129 | 1.6025 |
| d1ZnO | 1.9512 | 1.9920 |
| d2ZnO | 1.9791 | 2.0120 |
| dZnTi | - | 3.3464 |
| dTiO | - | 1.8391 |
| Energy (a.u) | -2757.3383 | -2755.1890 |

The inter-atomic distances in Table 6 provide detailed insights into how different configurations of Ti doping affect the atomic arrangements and bonding energies within ZnO. Configuration (g) and configuration (h) show consistent patterns in bond lengths, whereas configuration (i) exhibits variations, particularly in dTi-O and dTi-Zn, reflecting the influence of Ti concentration on local structural properties. The concentration of Ti in the crystal has a significant impact on its stability. As shown in Table 7, the total energy increased by approximately ~136 eV at 5.55 % Ti and by about ~190 eV at 8.33 %. There is also a slight modification in the total energy between configurations (h) and (g), indicating that the position of Ti in the crystal does not have a significant impact.

*Table 7: Inter-atomic distances and total energy for different configurations of ZnO doped with Ti.*

|     | d1Zn-O (Å) | d2Zn-O (Å) | dTi-O (Å) | dTi-Zn (Å) | E (eV)          |
| --- | ---------- | ---------- | --------- | ---------- | --------------- |
| (g) | 2.000      | 1.998      | 2.007     | 3.110      | -2752.8772189094 |
| (h) | 1.999      | 1.995      | 1.946     | 3.249      | -2752.8966584565 |
| *(i)* | 2.003    | 2.018      | 2.018     | 3.193      | -2750.6078872289 |

### 3.3.2 Electronic Properties

The ZnO electronic properties are significantly influenced by doping with Titanium (Ti), which alters its band structure and optoelectronic characteristics [42,43]. In this study, we examined several configurations of Ti-doped ZnO, corresponding to 2.77%, 5.55%, and 8.33%. As detailed in Table 8, our investigation revealed band gap energies ($E$g) of 2.001 eV, 2.131 eV, 2.241 eV, and 2.400 eV, respectively. This decrease in behavior of the band gap was reported and discussed by several theoretical studies [43]. This is probably due to the oxygen vacancy, as detected by the *ESR* measurement, which may act as an electron acceptor below the ZnO conduction band influencing therefore the band structure. The Quantum confinement effect can also modify the band structures of configurations (a), (b), (c) and (d). These configurations depicted in Figure 8, illustrate that the system turns from a direct band gap semiconductor into an indirect band gap. This kind of transformation was reported by several studies but with other doping elements such as carbon [43].

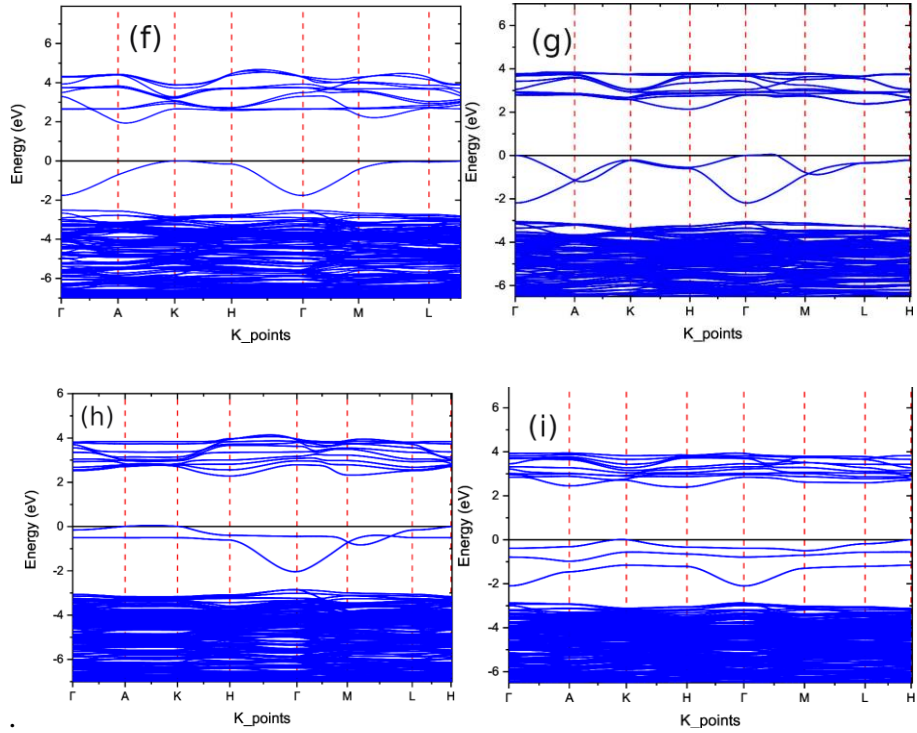

*Fig. 8. Band structure of ZnO for the configurations (f), (g), (h) and (i) doped with different concentration (2.77%, 5.55%, 5.55%, and 8.33%) of Ti, Fermi level referenced to 0.*

To illustrate the contributions of each electron orbital to the energy band, we have exhibited the DOS and PDOS of ZnO doped with 8.33% of Ti illustrated in configuration (i) shown in Figure 9. The two zones of the valence band, from -5 eV to -1 eV and from -9 eV to -3 eV, are derived from the O(2p) state and Zn(d) state, respectively. The two zones of the conduction band, from -1 eV to 0 eV and from 2.4 eV to 5 eV, are derived from the Ti(d) state and Zn(p) state, respectively. These observations explain the difference in the band gap between the pure supercell and the one doped with 8.33% of Ti. The PDOS for each element given in Figure 9, show the contributions of the different electronic states. The observed decrease in the band gap with increasing Ti concentration (from 2.77% to 8.33%) aligns well with theoretical predictions and experimental findings [45,46]. Previous studies have reported similar trends, indicating that higher Ti concentrations induce a narrowing of the band gap due to the introduction of additional electronic states within the band structure.

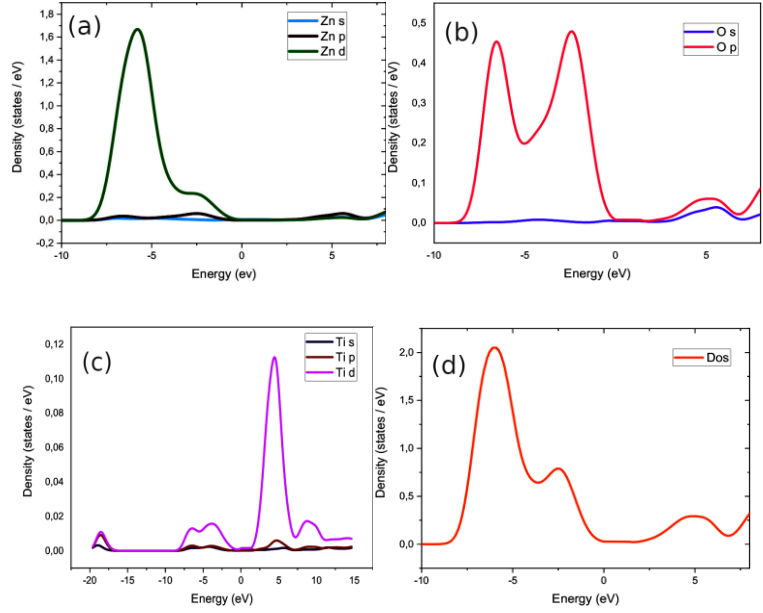

*Fig. 9. DOS and PDOS for configuration (i) corresponding to $8.33\%$ of Ti. Panels (a), (b), and (c) show the contributions from Zn, O, and Ti atoms respectively, while panel (d) displays the total density of states. The Fermi level is referenced to 0 eV.*

Table 8: Calculated Mg-ZnO band gap energies.

| Configuration | This Work Eg | |
| --- | --- | --- |
| | DFT | DFT+U |
| (f) | 0.371 | 2.001 |
| (g) | 0.601 | 2.231 |
| (h) | 0.651 | 2.281 |
| (i) | 0.769 | 2.399 |

In conclusion, the structural analysis reveals that Ti doping causes an expansion in ZnO's lattice parameters and modifies interatomic distances, leading to significant changes in the crystal structure. Higher Ti concentrations increase the total energy, indicating a reduction in stability. Additionally, Ti doping shifts ZnO from a direct to an indirect band gap semiconductor, with a decrease in the band gap energies. This trend aligns with theoretical and experimental findings, highlighting Ti's influence on ZnO's electronic properties. The observed modifications in lattice constants and bond lengths suggest potential applications in optoelectronics and semiconductor technologies. Future studies could explore the impact of varying doping levels and positions on other properties. To maintain ZnO structure stability like that found with Mg doping; while improving electrical properties, we tried co-doping ZnO with Mg and Ti.

### 3.4. ZnO Co-doping with Magnesium (Mg), and Titanium (Ti)

### 3.4.1 Structural Properties Optimization

After studying the geometrical and electronic properties of the pure ZnO supercell and investigating the impact of doping with different concentrations of Ti and Mg, we tried to explore the effect of co-doping the pure cell with 5.54%, 11.10% and 16.66% of Ti and Mg each in three different configurations.

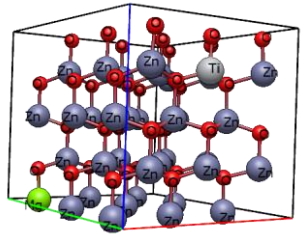

*Fig. 10. ZnO for the configuration (j) co-doped with one atom of Mg and one atom of Ti.*

We present in Table 9 the structural parameters of ZnO co-doped with Ti and Mg each with a proportion of 5.54%. These parameters are derived from geometric optimization resulting in configuration (j). This one is depicted in Figure 10. Table 9 shows that the lattice constants $a$ and $c$ present a marginal decrease and increase, respectively, indicating a minor distortion in the ZnO lattice. The (c/a) ratio decreases, suggesting a slight elongation of the unit cell along the c-axis. The total energy of the supercell in configuration (j) is closer to that of the pure one more than the other doped configurations with Ti or Mg, indicating a promising compound which combines the geometry and stability of pure ZnO with new electronic properties introduced by the presence of impurities.

*Table 9: ZnO in Wurtzite geometry (72 atoms) co-doped with one of Mg and one atom of Ti.*

| Parameter | This work ZnO | This work Zn-TiMgO |
|---|---|---|
| $a$ (Å) | 3.2375 | 3.2200 |
| $c$ (Å) | 5.2220 | 5.4400 |
| $c/a$ | 1.6129 | 1.5174 |
| $d1ZnO$ (Å) | 1.9512 | 1.9920 |
| $d2ZnO$ | 1.9791 | 2.0120 |
| $dZnTi$ | - | 3.2300 |
| $dTiO$ | - | 1.8391 |
| Total Energy (a.u) | -2757.3383 | -2758.0374 |

In order to understand the impact of the number and position of impurties on the stability, geometry and electron properties of $Zn_{1-x-y}Mg_xTi_yO$, we optimize two other configurations (k) and (l). The inter-atomic distances and total energy for different configurations of ZnO co-doped with Mg and Ti are listed in Table 10. The codoping introduces significant changes in the structural parameters compared to the pure and singly doped ZnO.

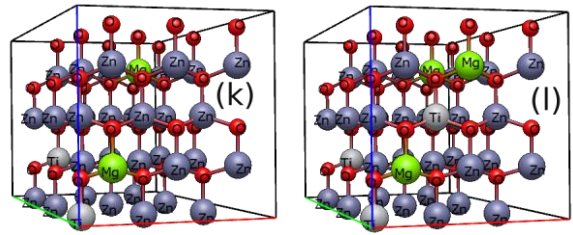

Fig. 11. Configurations (k) and (l) of ZnO doped with 5.55% and 8.33% of Mg.

Table 10: Inter-atomic distance and total energy for different configurations of ZnO co-doped with Mg and Ti.

| | $d1Zn-O$ | $d2Zn-O$ | $dMg-O$ | $dMg-Zn$ | $dTi-O$ | $dTi-Zn$ | $E$ (a.u) |
|---|---|---|---|---|---|---|---|
| (k) | 2.000 | 1.998 | 2.007 | 3.180 | 2.001 | 3.110 | -2756.5171 |
| (i) | 1.999 | 1.994 | 2.025 | 3.243 | 2.007 | 3.110 | -2757.8271 |

The total energy calculated for the co-doped configurations (j), (k) and (i) are close to that of the pure supercell. This confirms the stability of $Zn_{1-x-y}Mg_xTi_yO$ and the importance of more investigating its properties. The changes in bond lengths and lattice parameters can be correlated with the differences in ionic radii and bonding characteristics of the dopant atoms. Ti has a larger ionic radius than Zn, leading to an expansion of the ZnO lattice when it is introduced. Conversely, Mg has a smaller ionic radius, which can cause a contraction in the lattice. The combined effect of both dopants results in the observed structural modifications. These findings are consistent with previous studies that have reported similar trends in structural changes upon co-doping ZnO with different elements. For instance, Li et al. [47] observed that co-doping ZnO with *Ga* and *N* resulted in significant modifications of the lattice parameters and inter-atomic distances, similar to our observations for *Mg* and *Ti* co-doping. Additionally, Xu et al. [48,49] highlighted that the stability and electronic properties of co-doped ZnO are highly dependent on the specific arrangement and concentration of the dopants.

### 3.4.2 Electronic Properties

By analyzing the electronic properties of ZnO co-doped with Ti and Mg at different concentrations with a band structure calculation, it is revealed that doping ZnO with Ti and Mg influences significantly its electronic properties. In fact, the band structure obtained for various configurations (j), (k) and (l)) depicted in Figure 12, show that the $Zn_{1-x-y}Mg_xTi_yO$ has an indirect band gap. This caracter is due to the presence of Titanium. As shown in Table 10, $E$g decreases by ~35 % compared to that of the pure ZnO.

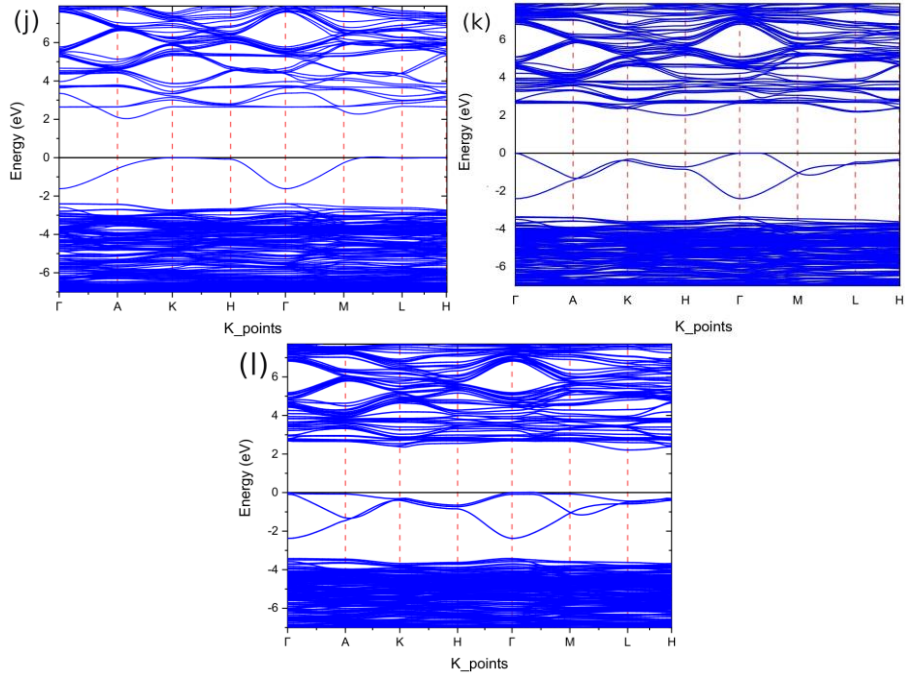

*Fig. 12. Band structure for configuration (j), (k) and (l) corresponding to ZnO co-doped with 5.54% (panel a), 11.10% (panel b) and 16.66% (panel c) of Mg and Ti. The Fermi level is referenced to 0 eV.*

*Table 11: Band gap energies of ZnO configurations co-doped with Mg and Ti.*

| Configuration | This Work Eg |
| --- | --- |

|     | DFT   | DFT+U |
|-----|-------|-------|
| (j) | 0.471 | 2.101 |
| (k) | 0.373 | 2.003 |
| (m) | 0.580 | 2.210 |

To illustrate the contributions of each electron orbital to the energy band, we have exhibited the pure ZnO DOS together with the PDOS of zinc, oxygen, magnesium and titanium in Figure 13. The DOS and PDOS are plotted from -10 to 8 eV with the Fermi level referenced to 0 eV and a gap of 2.4 eV. The two zones of the valence band, from -9 eV to -4 eV and from -9 eV to 3 eV, are derived from the O(2p) state and Zn(d) state, respectively. The two zones of the conduction band, from -4 eV to 0 eV and from -2.4 eV to 5 eV, are derived from the O(p) state and Ti(d) state, respectively. Additionally, the zones from 2 eV to -4 eV and -2.4 eV to 5 eV are derived from the Ti(d) state and Mg(p) state, respectively. Also, the zones from 4 eV to -10 eV and -2.4 eV to 5 eV are derived from the Mg(p) state and O(p) state, respectively. These observations explain the difference of 1.2 eV (as shown in Table 11) in the band gap between the pure supercell and the one co-doped with 16.66% of Ti and Mg. The introduction of Ti and Mg atoms into the ZnO lattice leads to perturbation in the electronic states, which modifies the band structure as shown on panels (a), (b), (c) and (d) of Figure 12. For configuration (j) with 5.54% doping, the band gap of 2.10 eV is lower than the band gap of pure ZnO, which is typically around 3.4 eV at room temperature [4]. This band gap reduction can be explained by the introduction of impurity states within the band gap due to the dopants, which narrows the effective band gap as in the case of configurations (f), (j), (h) and (i) of ZnO doped with Ti and configurations (j), (k) and (l) of ZnO co-doped with Mg and Ti. In configuration (k) with 11.10% doping, the band gap further decreases to 2.00 eV. This trend suggests that increasing the concentration of dopants introduces more impurity levels, which further disrupts the conduction and valence bands, leading to a narrower band gap. Interestingly, for configuration (m) with 16.66% doping, the band gap increases to 2.21 eV. This anomaly can be interpreted as a result of the complex interplay between the Ti and Mg dopants at higher concentrations. At this concentration, the dopants might interact in such a way that the resulting electronic structure leads to a slight widening of the band gap compared to lower doping levels.

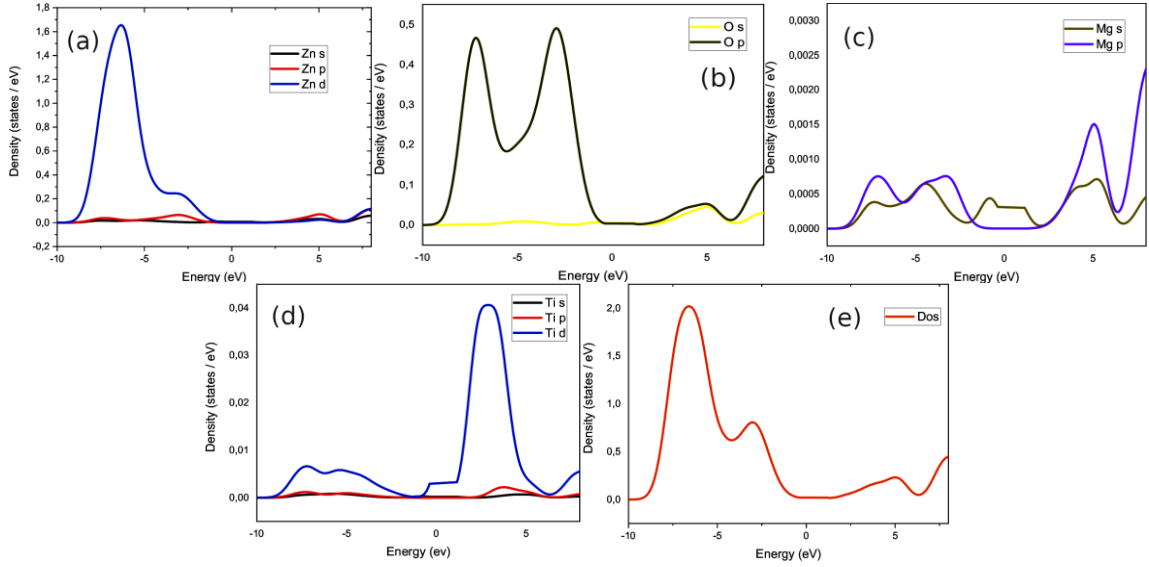

*Fig. 13. DOS and PDOS for configuration (l) correspond to co-doped with three atoms of Mg and three atoms Ti. Panels (a), (b), (c) and (d) show the contributions from Zn, O, Ti and Mg atoms respectively, while panel (e) displays the total density of states, and the Fermi level has been referenced to 0 eV.*

Thus, the co-doping of ZnO with Ti and Mg at different concentrations leads to significant modifications in its electronic properties, particularly the band gap. The observed trends in band gap energies provide insights into the complex interactions between the dopants and the ZnO lattice, which could be useful for tuning the material's properties for specific applications in optoelectronic devices.

## 4. Conclusion

In summary, the structural and electronic properties of (Mg and Ti) doping and codoping $Zn_{1-x-y}Mg_xTi_yO$ system has been studied based on DFT calculations. By systematically varying the concentrations and substitution positions of Mg and Ti atoms, we identified optimal configurations that closely matched experimental data. Our findings demonstrate that co-doped ZnO could serve as a promising material for applications requiring enhanced optical and electronic properties. Future research may focus on further optimizing doping strategies and exploring additional dopants to tailor the properties of ZnO for specific technological applications in optoelectronic devices.

**Acknowledgement**

We would like to thank Prof. Habib Bouchriha from University of Tunis El Manar for his useful discussions.

## Authors Contributions

*Sidi Ahmedbowba*: Visualization, Numerical Simulation, and Writing. *Fehmi Khadri*: Visualization, Simulation, Validation, Analysis and Writing. *Walid Ouerghui*: Validation of the results. *Said Ridene*: Conceptualization, Supervision, Validation, Writing and Correction.

## Declarations

**Data Availability Statement:** No Data associated in the manuscript.

**Conflict of interest:** The authors declare that they have no known competing financial interests or personal relationships that could have appeared to influence the work reported in this paper.

## References

[1] Iwan A, Tazbir I, Sibinski M, Boharewicz B, Pasciak G, Schab-Balcerzak E, Mater. Sci. Semicond. Process., 2014, 24, 110.
[2] Chen M, Pei ZL, Sun C, Gong J, Huang RF, Wen LS, Mater. Sci. Eng. B, 2001, 85, 212.
[3] Look DC, Mater. Sci. Eng. B, 2001, 80, 383.
[4] Yu CF, Sun SJ, Hsu HS, Phys. Lett. A, 2015, 379, 211.
[5] Ryu YR, Lee TS, Lubguban JA, White HW, Park YS, Youn CJ, Appl. Phys. Lett., 2005, 87, 153504.
[6] Ozgür. U et al., J. Appl. Phys., 2005, 98, 041301.
[7] Wu HY, Cheng XL, Hu CH, Zhou P, Physica B, 2010, 405, 606.
[8] Park WI, Yi GC, Jang HM, Appl. Phys. Lett., 2001, 79, 2022.
[9] Vashaei Z, Minegishi T, Suzuki H, Hanada T, Cho MW, Yao T, Center for Interdisciplinary Research, Tohoku University, Aramaki-Aza-Aoba, Aoba-ku, Sendai 980-8578, Japan.
[10] Yamamoto T, Katayama-Yoshida H, Jpn. J. Appl. Phys., 1999, 38, L166.
[11] P. Raji, K. Balachandra Kumar, Investigation of Ti doping on the structural, optical and magnetic properties of ZnO nanoparticles.
[12] N. Kılınç, L. Arda, S. Öztürk, Z.Z. Öztürk, Cryst. Res. Technol., 2010, 45, 529.
[13] Ma X, Wu Y, Lv Y, Zhu Y, J. Phys. Chem. C, 2013, 117(49), 26029-26039.
[14] Uddin J, Scuseria GE, Phys. Rev. B, 2006, 74, 245115.
[15] Baym G, Kadanoff LP, Phys. Rev., 1961, 124, 2879.
[16] Tran F, Blaha P, Phys. Rev. Lett., 2009, 102, 226401.
[17] Krack M, Parrinello M, J. Grotendorst (Ed.), High Performance Computing in Chemistry, IC-Directors, 2004, 25, pp. 29.
[18] Perdew JP, Ruzsinszky A, Csonka GI, Vydrov OA, Scuseria GE, Constantin LA, Zhou X, Burke K, Phys. Rev. Lett., 2009, 102, 039902.
[19] N. Hamzah, M.H. Samat, N.A. Johari, A.F.A. Faizal, O.H. Hassan, A.M.M. Ali, R. Zakaria, N.H. Hussin, M.Z.A. Yahya, M.F.M. Taib, First-principle LDA+U and GGA+U calculations on structural and electronic properties of wurtzite ZnO.
[20] A.A. Mohamad, M.S. Hassan, M.K. Yaakob, M.F.M. Taib, F.W. Badrudin, O. H. Hassan, M.Z.A. Yahya, First-principles calculation on electronic properties of zinc oxide by zincir system, J. King Saud Univ. - Eng. Sci. 29 (2017) 278-283.
[21] K. Harun, N.A. Salleh, B. Deghfel, M.K. Yaakob, A.A. Mohamad, DFT+U calculations for electronic, structural, and optical properties of ZnO wurtzite structure: A review, Results Phys. 16 (2020) 102829.
[22] S.J. Clark, M.D. Segall, C.J. Pickard, P.J. Hasnip, M.I. Probert, K. Refson, et al., First principles methods using CASTEP, Z. Kristallogr. 220 (2005) 567-570.
[23] F. Oba, M. Choi, A. Togo, I. Tanaka, Point defects in ZnO: an approach from first principles, Sci. Technol. Adv. Mater. 12 (2011) 034302.
[24] W. K\"orner, C. Els\"asser, First-principles density functional study of dopant elements at grain boundaries in ZnO, Phys. Rev. B 81 (2010) 085324.
[25] B. Himmetoglu, A. Floris, S. Gironcoli, M. Cococcioni, Hubbard-corrected DFT energy functionals: the LDA+U description of correlated systems, Int. J. Quantum Chem. 114 (2014) 14-49.
[26] V.I. Anisimov, J. Zaanen, O.K. Andersen, Band theory and Mott insulators: Hubbard U instead of Stoner I, Phys. Rev. B 44 (1991) 943.

[27]     X.Y. Deng, G.H. Liu, X.P. Jing, G.S. Tian, On-site correlation of p-electron in $d^{10}$ semiconductor zinc oxide, Int. J. Quantum Chem. 114 (2014) 468-472.
[28]     X. Ma, Y. Wu, Y. Lv, Y. Zhu, Correlation effects on lattice relaxation and electronic structure of ZnO within the GGA+U formalism, J. Phys. Chem. C 117 (2013) 26029-26039.
[29]     Desgreniers S, High-density phases of ZnO: Structural and compressive parameters, Phys. Rev. B, 1998, 58(21), 14102-14105.
[30]     Zhang YG, Zhang GB, Xu WY, First-principles study of the electronic structure and optical properties of Ce-doped ZnO, J. Appl. Phys., 2011, 109, 063510.
[31]     Schleife A, Fuchs F, Furthmler J, Bechstedt F, First-principles study of ground- and excited-state properties of MgO, ZnO, and CdO polymorphs, Phys. Rev. B, 2006, 73(24), 245212.
[32]     Ahmoum H, Boughrara M, Su'ait MS, Chopra S, Kerouad M, Impact of position and concentration of sodium on the photovoltaic properties of zinc oxide solar cells, Physica B, 2019, 560, 28-36.
[33]     Reynolds DC, Look DC, Jogai B, Litton CW, Cantwell G, Harsch WC, Valence-band ordering in ZnO, Phys. Rev. B, 1999, 60(4), 2340-2344.
[34]     M. Yaakob, N. Hussin, M. Taib, T. Kudin, O. Hassan, A. Ali, et al., First principles LDA+U calculations for ZnO materials, Integr. Ferroelectr. 155 (2014) 15-22.
[35]     Y.-S. Lee, Y.-C. Peng, J.-H. Lu, Y.-R. Zhu, H.-C. Wu, Electronic and optical properties of Ga-doped ZnO, Thin Solid Films 570 (2014) 464-470.
[36]     M.H. Huang, S. Mao, H. Feick, H. Yan, Y. Wu, H. Kind, et al., Room-temperature ultraviolet nanowire nanolasers, Science 292 (2001) 1897-1899.
[37]     C. Dong, C. Persson, L. Vayssieres, A. Augustsson, T. Schmitt, M. Mattesini, et al., Electronic structure of nanostructured ZnO from x-ray absorption and emission spectroscopy and the local density approximation, Phys. Rev. B 70 (2004) 195325.
[38]     Rouchdi M, Salmani E, Fares B, Hassanain N, Mzerd A, Synthesis and characterics of Mg doped ZnO thin films: Experimental and ab-initio study, Results Phys., 2017, 7, 620-627.
[39]     Bilgili O, The Effects of Mn Doping on the Structural and Optical Properties of ZnO, Acta Phys. Pol. A, 2019, 136, 460-466.
[40]     Fang D, Li C, Wang N, Li P, Yao P, Structural and optical properties of Mg-doped ZnO thin films prepared by a modified Pechini method, Cryst. Res. Technol., 2013, 48(5), 265-272.
[41]     K.R. Murali, Structural, Optical and Electrical Properties of Spray Pyrolysed Ti-Doped ZnO Films, ECS Meeting Abstracts MA2014-02 (43) p. 2077 (2014).
[42]     T. Munir, M. Kashif, W. Hussain, A. Shahzad, M. Imran, A. Ahmed, N. Amin, N. Ahmed, A. Hussain, M. Noreen, First principles study of structural and electronic properties of Ti doped ZnO, Physics Department, Government College University, Faisalabad, Pakistan.
[43]     P. Si, X. Su, Q. Hou, Y. Li, W. Cheng, First-principles calculation of the electronic band of ZnO doped with C, J. Semicond. 30 (2009) 052001.
[44]     L.A. Xue, Y. Chen, R.J. Brook, The Influence of Ionic Radii on the Incorporation of Trivalent Dopants into BaTiO\(_3\), Dep. Ceram., Univ. Leeds, U.K. (1988).
[45]     Samuel J, Suresh S, Shabna S, Sherlin Vinita V, Joslin Ananth N, Shajin Shinu PM, Mariappan A, Turibius Simon, Samson Y, Biju CS, Characterization and antibacterial activity of Ti doped ZnO nanorods prepared by hydrazine assisted wet chemical route, Physica E, 2022, 143, 115374.
[46]     Ma L, Ai X, Huang X, Ma S, Effects of the substrate and oxygen partial pressure on the microstructures and optical properties of Ti-doped ZnO thin films, Superlattices Microstruct., 2011, 50(6), 703-712.
[47]     Li H, Lv Y, Li J, Yu K, First-principles study of p-type conductivity of N-Al/Ga/In co-doped ZnO, Phys. Scr., 2015, 90(2), 025803.
[48]     Xu L, Li X, Chen Y, Xu F, Structural and optical properties of ZnO thin films prepared by sol-gel method with different thickness, Appl. Surf. Sci., 2011, 257(9), 4031-4037.
[49]     D. Cherrad, First-principles studies on (001) surface electronic bonding and magnetic properties of ZnCMn3 and ZnNMn3 intermetallic antiperovskites type compounds, J. Alloys Compd. 586 (2014) 230-238.